\documentclass[showpacs,twocolumn,aps,pra,superscriptaddress]{revtex4-1}
\usepackage{amssymb}
\usepackage{amsmath}
\usepackage{epsfig}
\begin{document}

\title{Mesoscopic Mechanical Resonators as Quantum Non-Inertial Reference Frames}

\author{B. N. Katz}
\altaffiliation{Present address: Department of Physics, Pennsylvania State University, University Park, Pennsylvania 16802, USA}
\author{M. P. Blencowe}
\affiliation{Department of Physics and Astronomy, Dartmouth College, New Hampshire 03755, USA}
\author{K. C. Schwab}
\affiliation{Applied Physics, California Institute of Technology, Pasadena, CA 91125, USA}
\affiliation{Kavli Nanoscience Institute, Pasadena, CA 91125, USA}

\date{\today}

\begin{abstract}
An atom attached to a micrometer-scale wire that is vibrating at a frequency $\sim 100~{\mathrm{MHz}}$ and with displacement amplitude~$\sim 1~{\mathrm{nm}}$ experiences an acceleration magnitude $\sim 10^9~{\mathrm{ms}}^{-2}$, approaching the surface gravity of a neutron star. As one application of such extreme non-inertial forces in a mesoscopic setting, we consider a model two-path atom interferometer with one path consisting of the $100~{\mathrm{MHz}}$ vibrating wire atom guide. The vibrating wire guide serves as a non-inertial reference frame and induces an in principle measurable phase shift in the wave function of an atom traversing the wire frame. We furthermore consider the effect on the two-path atom wave interference when the vibrating wire is modeled as a quantum object, hence functioning as a quantum non-inertial reference frame. We outline a possible realization of the vibrating wire, atom interferometer using a superfluid helium quantum interference setup.     
\end{abstract}


\maketitle

\section{Introduction}

During the past decade there has been a growing effort to demonstrate nano-to-mesoscale mechanical systems existing in manifest quantum states~\cite{blencowe04,schwab05,poot12,chen13,aspelmeyer13}. One motivation is to understand how classical dynamics arises from quantum dynamics for systems with center of mass much larger than that of a single atom, and in particular establish whether the quantum-to-classical transition is solely a consequence of environmentally induced decoherence~\cite{zurek91} or perhaps ultimately due to some as yet undiscovered, fundamental `collapse' mechanism~\cite{bassi13}. Three important  milestones have been the demonstration of a $\sim 5~{\mathrm{GHz}}$ mechanical resonator mode in a  single  phonon Fock state~\cite{oconnell10}, the demonstration of an entangled  $\sim 10~{\mathrm{MHz}}$ mechanical resonator mode--microwave cavity mode state~\cite{palomaki13}, and the demonstration of a $\sim 4~{\mathrm{MHz}}$ mechanical resonator in a quadrature-squeezed state with minimum variance $0.80$ times that of the quantum ground state~\cite{wollman15}.     

A particularly straightforward mechanical geometry is that of a long, thin suspended beam (wire) that is supported at both ends (i.e., doubly clamped). The wire can be driven transversely, exciting its fundamental flexural mode resonance, using several available actuation methods. For the example of  a crystalline silicon (Si) wire that is a few micrometers long and a fraction of a micrometer in cross section, the mechanical fundamental flexural frequency is $\Omega\sim 2\pi\times100~{\mathrm{MHz}}$~\cite{cleland96}. Consider an Si atom or other atom type attached to the surface midway along the length of such a vibrating wire. Suppose that the midpoint displacement amplitude is $X_0\sim 1~{\mathrm{nm}}$. Then assuming that the midpoint undergoes simple harmonic motion, $X(t)=X_0 \cos\left(\Omega t+\varphi_0\right)$, we have for the maximum acceleration experienced by the attached atom $\ddot{X}_{\mathrm{max}} =\Omega^2 X_0\sim 10^9~{\mathrm{ms}}^{-2}$.  This is an unexpectedly large acceleration, $10^8$ times larger than the gravitational acceleration $g$ on the surface of the Earth and closer in magnitude to the surface gravity of a neutron star~ \cite{bejger04}.

\begin{figure}[htbm]
\centering
{\includegraphics[width=6.0cm]{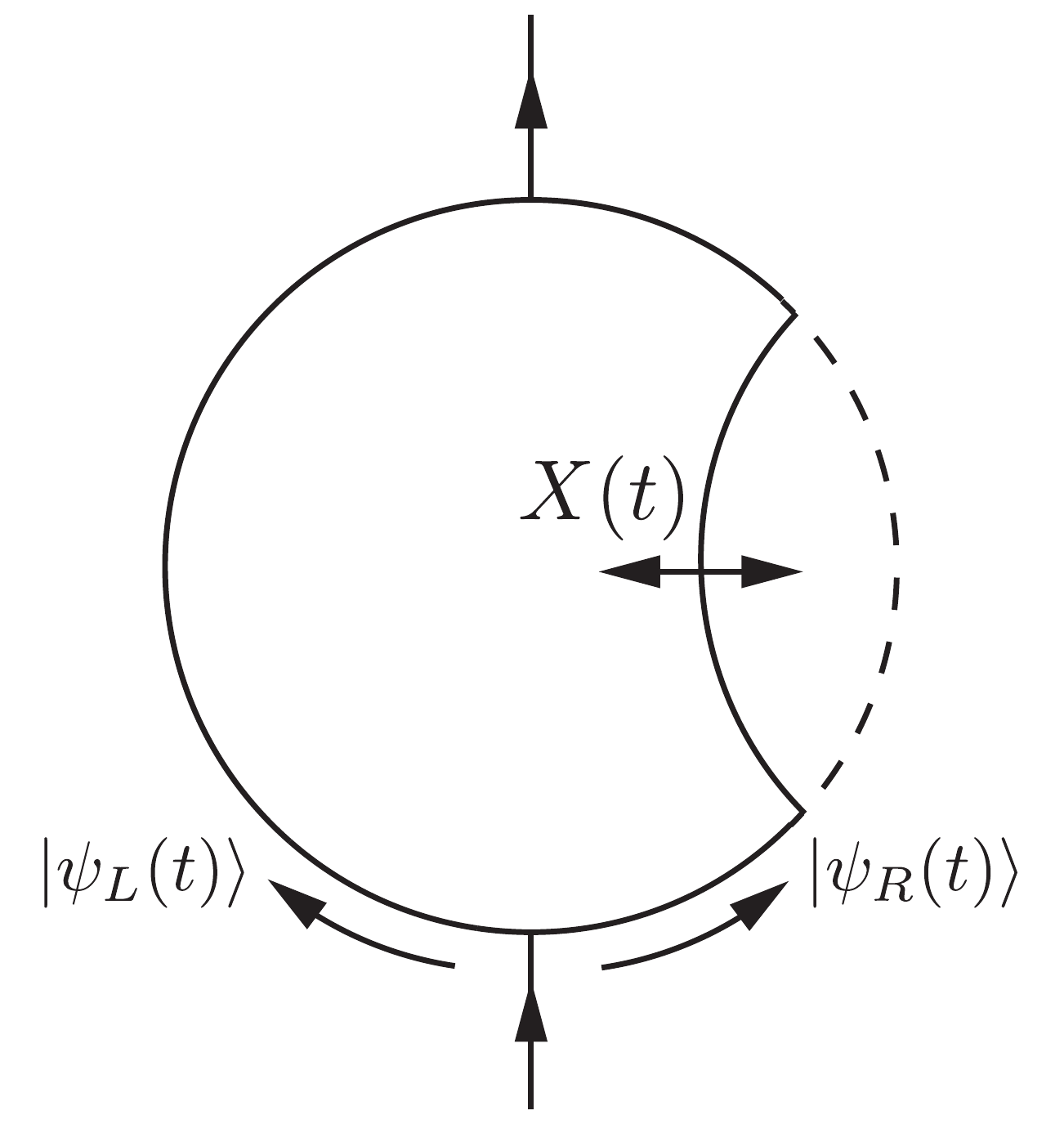}
}
\caption{Scheme of the two-path atom interferometer with non-inertial, vibrating wire frame forming part of the right path.}
\label{fig:scheme}
\end{figure}

In this paper, we analyze one possible application of these extreme accelerations in a mesoscopic setting; there are undoubtedly other hitherto unexplored applications. In particular, we shall consider a model two-path atom interferometer shown schematically in Fig.~\ref{fig:scheme}. The right path consists of a vibrating wire, atom guide segment with micrometer dimensions similar to those described above. The left path is fixed, without a vibrating wire segment. Incident atom wave packets split into left and right wave packet components. The right wave component will accumulate a phase shift relative to the left wave component as a consequence of the vibrating wire functioning effectively as a non-inertial reference frame for the traversing right wave. The right wave component eventually exits the non-inertial frame and recombines with the left wave component.  As we shall see below in Sec.~\ref{sec:classical}, this results in an in principle detectable fringe visibility for the left-right wave component interference as the vibrating wire amplitude is varied in the nanometer range. 

Quantum wave interference due to gravitational and inertial forces has been experimentally demonstrated for neutrons~\cite{colella75,werner79,bonse83,atwood84}, atoms~\cite{riehle91}, Cooper-pairs~\cite{zimmerman65}, and electrons~\cite{hasselbach93}, verifying the equivalence between these forces for quantum matter systems~\cite{greenberger83}. Furthermore, a recent comprehensive analysis takes into account also the possibility of atoms in each path of an interferometer experiencing different inertial forces, similar to our vibrating wire, interferometer model~\cite{roura14}. However, in all of these experiments and analyses the non-inertial reference frames (and of course gravity)  were treated as classical systems and there was no reason to view them otherwise. On the other hand, with mesoscopic mechanical resonators now being prepared and measured in manifest quantum states~\cite{oconnell10,palomaki13} as described above, it is very natural to consider the consequences for the matter wave interference of the vibrating wire functioning effectively as a quantum non-inertial reference frame~\cite{aharonov84,bartlett07,angelo11,angelo12,pereira15}. We shall therefore consider in Sec.~\ref{sec:quantum} the effects on the fringe visibility for the left-right wave component interference when the vibrating wire is described quantum-mechanically; the details of the calculations are given in the Appendix. Sec.~\ref{sec:realization} outlines a possible way to realize a non-inertial, vibrating wire-atom interferometer using a superfluid helium quantum interference setup. 

\section{classical frames}
\label{sec:classical}

Beginning first with a classical description, we approximate the non-inertial frame as a long, thin beam  (wire) of length $L$ with hinged boundary conditions, so that for small amplitude transverse displacements ($X_0\ll L$ ) the  frame coordinates in the lowest, fundamental flexural mode are $X(z,t)=X_0 \sin\left(\pi z/L\right)\cos\left(\Omega t+\varphi_0\right)$, $0\leq z\leq L$. We suppose that atoms traversing the frame  in the longitudinal, $z$ coordinate direction are described by localized wave packets propagating with uniform group velocity $v=L/T$ relative to the vibrating frame,  where $T$ is the atom dwell time in the frame. The atoms are  assumed to be confined by a harmonic potential to the frame in their transverse, $x$ coordinate direction that is aligned with the frame flexing $X$ coordinate direction. The bound atom potential is then $V(x,t)=\frac{1}{2}m\omega^2\left[x-X_0\sin \left(\pi t/T\right)\cos\left(\Omega t+\varphi_0\right)\right]^2$, where both atom and frame coordinates ($x$ and $X$, respectively) are defined relative to a common origin in the assumed inertial lab frame.   The various characteristic frequencies are assumed to satisfy $\pi/T \ll\Omega\ll\omega$, so that the atom spends many oscillation cycles in the frame, while it is tightly bound with negligible transverse motion relative to the frame. This assumption is not fundamental but rather for calculational convenience. In particular, assuming instead $\Omega\lesssim\omega$ requires a more involved analysis of the interference between the left and right waves, but can still result in significant accumulated phase differences.

Micrometer scale resonators have masses  $M\sim 10^{-15}~{\mathrm{kg}}$, while the atom mass $m\sim 10^{-27}~{\mathrm{kg}}$, i.e.,  $M\ggg m$. It is therefore reasonable to neglect the back action of the atom on the classical frame. We assume that the frame is initially excited in its fundamental transverse flexural mode and freely oscillates with negligible change in amplitude (i.e., damping) over the atom dwell time $T$. In terms of the transverse atom  $x$ and frame $X$  coordinates, the Schr\"{o}dinger equation for the atom in the vibrating frame described by the right path wave component $\psi_R(x,t)$ is then
\begin{eqnarray}
&&i\hbar\frac{\partial\psi_R}{\partial t}=-\frac{\hbar^2}{2m}\frac{\partial^2\psi_R}{\partial x^2}\cr
&&+\frac{1}{2}m\omega^2\left[x-X_0\sin \left(\pi t/T\right)\cos\left(\Omega t+\varphi_0\right)\right]^2\psi_R.
\label{sch1eq}
\end{eqnarray}      
This equation may be straightforwardly solved assuming a Gaussian function form. With the atom entering the frame initially in its transverse ground state: $\psi_R(x,0)=\left(\frac{m\omega}{\pi\hbar}\right)^{1/4}\exp\left(-\frac{m\omega x^2}{2\hbar}\right)$, the resulting interference between the left and right path wave components at $t=T$ is
\begin{equation}
\psi_L^*(x,T)\psi_R(x,T)=\sqrt{\frac{m\omega}{\pi\hbar}}e^{-\frac{m\omega x^2}{\hbar}+i\phi},
\label{interfere1eq}
\end{equation}
where the accumulated phase difference between the left and right waves is 
\begin{equation}
\phi\approx\frac{m\Omega^2 X_0^2 T}{8\hbar}\equiv\frac{m\Omega^2 X_0^2 L}{8\hbar v}.
\label{phase1eq}
\end{equation}
Note that the only atom attributes that the phase difference~(\ref{phase1eq}) depends on are its inertial mass $m$ and traversal velocity $v$ relative to the vibrating frame; $\omega$ does not appear to leading order as a consequence of the atom assumed to be tightly bound transversely to the frame, i.e., $\omega\gg\Omega$. Putting in some numbers: with $m\sim 10^{-27}~{\mathrm{kg}}$, $\Omega\sim 2\pi\times 10^8~{\mathrm{s}}^{-1}$, and $X_0\sim 10^{-9}~{\mathrm{m}}$, we have $\phi\sim 0.5~T(\mu{\mathrm{sec}})$. Thus, significant phase shifts result for dwell times in excess of a microsecond.

Interestingly, this estimated phase shift  following from Eq.~(\ref{phase1eq}) is of the same order of magnitude as the measured  gravitational phase difference between two interfering neutron beam paths in Ref.~\cite{colella75}, despite the fact that the vibrating wire acceleration magnitude $\Omega^2 X_0$ is eight orders of magnitude larger than $g\approx 10~{\mathrm{m s}}^{-2}$. To understand this, consider the gravitational phase difference expression~\cite{greenberger83}: $\phi=m g_{\alpha} A/(\hbar v)$, where  $m$ is the neutron mass, $g_{\alpha}$ is the component of gravity in the plane of the neutron paths, $A$ is the area enclosed by the two paths, and $v$ is the neutron velocity.  In contrast to this expression, Eq.~(\ref{phase1eq}) does not scale with the area enclosed by the two paths. This is because the transverse inertial force experienced by  the atom is  different in the two paths, in particular much larger in the right path containing the vibrating wire, whereas the gravitational field is uniform across the neutron interferometer. Comparing Eq.~ (\ref{phase1eq}) with the gravitational phase difference expression, the reason for the similar phase shift magnitudes is that the effective area to atom velocity ratio term $X_0 L/v$ is about eight orders of magnitude smaller than the corresponding ratio $A/v$ for the neutron interferometer. However, in contrast to the much larger $\sim 10~{\mathrm{cm}}$ scale neutron interferometer where the acceleration due to gravity is of course well described classically, the micrometer scale, non-inertial vibrating wire frame can also in principle be prepared in a manifest quantum state.

A possibly more fundamental way to understand  the phase difference~(\ref{phase1eq}) follows from the original observation of de Broglie~\cite{debroglie} that the phase of a particle's wave function can be directly expressed in terms of the proper time along the path of the particle. In particular, we have~\cite{greenberger83,greenberger12}:
\begin{equation}
\phi =-\frac{m c^2}{\hbar}\left( \int_{t=0}^{t=T} d\tau_R- \int_{t=0}^{t=T} d\tau_L\right),
\label{phase2eq}
\end{equation}
where $c$ is the speed of light in vacuum and  $\tau_{L(R)}$ is the proper time elapsed for the atom traveling along the left (right) interferometer path. Since the frame velocity $\dot{X}_{\mathrm{max}} =\Omega X_0\sim 1~{\mathrm{ms}}^{-1}$ for the above considered parameters, we have  $|v_{L(R)}(t)|\ll c$ and hence $d\tau_{L(R)}=\sqrt{1-\left(v_{L(R)}(t)/c\right)^2}dt\approx\left(1-1/2 \left(v_{L(R)}(t)/c\right)^2\right)dt$, where $v_{L(R)}(t)$ is the left (right) path atom velocity relative the lab frame.  Substituting in the approximation $v_R(t)\approx -X_0\Omega\sin\left(\pi t/T\right)\sin\left(\Omega t+\varphi_0\right)$, valid for the condition $\pi/T \ll\Omega\ll\omega$, Eq.~(\ref{phase2eq}) then gives the same result as Eq.~(\ref{phase1eq}). Thus, the phase difference~(\ref{phase1eq}) can be viewed as a consequence of the ``twin paradox"~\cite{greenberger83,greenberger12}, where the right path wave packet bound to its noninertial, vibrating frame ``ages" less than the left path wave packet during the elapsed lab coordinate time interval $T$.

\section{quantum frames}
\label{sec:quantum}

Moving on now to treating the vibrating frame as a quantum system, the Schr\"{o}dinger equation for the composite atom-frame wave function $\Phi_R(x,X,t)$ is:
\begin{eqnarray}
&&i\hbar\frac{\partial\Phi_R}{\partial t}=-\frac{\hbar^2}{2 m} \frac{\partial^2\Phi_R}{\partial x^2}-\frac{\hbar^2}{2 M} \frac{\partial^2\Phi_R}{\partial X^2}\cr
&&+\frac{1}{2}M\Omega^2 X^2 \Phi_R +\frac{1}{2}m\omega^2\left[x-X\sin\left(\pi t/T\right)\right]^2 \Phi_R,
\label{sch2eq}
\end{eqnarray}
where we neglect the coupling between the frame and its dissipative environment and $M$ is the effective motional mass of the frame in the fundamental flexural mode.
The atom is assumed to enter the frame initially in its transverse ground state: $\psi_R(x,0)=\left(\frac{m\omega}{\pi\hbar}\right)^{1/4}\exp\left(-\frac{m\omega x^2}{2\hbar}\right)$, with $\Phi_R(x,X,0)=\psi_R(x,0)\Psi_R(X,0)$ for some initial prepared frame state $\Psi_R(X,0)$.

A potential puzzle concerns the fact that, as a quantum object,  the frame also has a phase and hence there may be an ambiguity concerning the part of the phase that ``belongs" to the atom. This puzzle is resolved by noting that only the atoms are detected at the interferometer output, so that the frame state must be traced over in the interference term, which can be expressed as (cf. the classical frame interference Eq.~(\ref{interfere1eq})) [see Appendix]:
\begin{eqnarray}
{\mathrm{Tr}}_{\mathrm{frame}}&&\left[\langle x|\hat{U}_R(T)|\psi_R(0)\rangle\langle\psi_L(0)|\otimes\hat{\rho}_{\mathrm{frame}}(0)\hat{U}_L^{\dagger}(T)|x\rangle\right]\cr
&&=\sqrt{\frac{m\omega}{\pi\hbar}}e^{-\frac{m\omega x^2}{\hbar}} \left\langle e^{i\hat{\phi}}\right\rangle.
\label{interfere2eq}
\end{eqnarray}
Here, we allow for the possibility that the frame is initially in a mixed state, while the unitary operators $\hat{U}_{L(R)}(T)$ implement the atom-frame evolution over the time interval $T$ when the atom is either attached ($R$) or not attached ($L$)  to the frame. Formally, we write the accumulated phase difference as an average, $ \left\langle e^{i\hat{\phi}}\right\rangle$, reflecting the fact that the frame is now in a quantum state.

Equations~(\ref{sch2eq}) and (\ref{interfere2eq}) may be straightforwardly solved by transforming (\ref{sch2eq}) to instantaneous normal mode coordinates such that (\ref{sch2eq}) becomes separable, and decomposing the initial frame state $\hat{\rho}_{\mathrm{frame}}(0)$ in terms of a coherent state basis $|\alpha\rangle$, where $\alpha=\sqrt{M\Omega/(2\hbar)} X+i P/\sqrt{2 M\hbar\Omega}$. Assuming Gaussian  solutions and taking advantage of the separation of frequency scales,  $\pi/T \ll\Omega\ll\omega$, we obtain (see Appendix for the details of the derivation):
\begin{equation}
 \left\langle e^{i\hat{\phi}}\right\rangle\approx\frac{1}{\pi}\int d^2\alpha\langle\alpha|\hat{\rho}_{\mathrm{frame}}(0)|\alpha e^{im\Omega T/(4M)}\rangle e^{i\frac{m}{8 M}(\Omega-\omega)T}.
 \label{interfere3eq}
 \end{equation}
 
 Consider the example of a frame initially in a displaced coherent state: $\hat{\rho}_{\mathrm{frame}}(0)=|\Psi_R(0)\rangle\langle\Psi_R(0)|=|\alpha_0\rangle\langle\alpha_0|$, where $\alpha_0=\sqrt{M\Omega/(2\hbar)} X(0) +iP(0)/\sqrt{2 M\hbar\Omega}$, with $X(0)=X_ 0 \cos\varphi_0$ and $P(0)=-M\Omega X_0\sin\varphi_0$. Equation~(\ref{interfere3eq}) then gives 
 \begin{eqnarray}
 \left\langle e^{i\hat{\phi}}\right\rangle&\approx&e^{|\alpha_0|^2 \left(e^{i\frac{m}{4M}\Omega T}-1\right)+i\frac{m}{8M}(\Omega-\omega)T}\cr
 &\approx& e^{i\frac{m}{4M}|\alpha_0|^2\Omega T+i\frac{m}{8M}(\Omega-\omega)T}\cr
 &=& e^{ i\frac{m\Omega^2 X_0^2 T}{8\hbar}+i\frac{m}{8M}(\Omega -\omega)T},
\label{phase3eq}
 \end{eqnarray}
where the approximation in the second line assumes that  $m\Omega T/(4 M)\ll 1$ and $|\alpha_0|^2 (m\Omega T/(4 M))^2\ll 1$. These conditions put an upper limit on the magnitude of the atom dwell time $T$ such that the  back action of the atom on the frame is negligible. Equation~(\ref{phase3eq})  coincides with the classical frame  accumulated phase difference~(\ref{phase1eq}), as we would expect for a coherent frame state. 

We now give two examples of manifest quantum frame states.  As our first example, consider a frame initially in a Fock state: $\hat{\rho}_{\mathrm{frame}}(0)=|N\rangle\langle N|$, $N=0, 1, 2,\dots$. Equation~(\ref{interfere3eq}) then gives 
\begin{equation}
 \left\langle e^{i\hat{\phi}}\right\rangle\approx e^{ i\frac{m}{4M}N\Omega T+i\frac{m}{8M}(\Omega-\omega)T}.
 \label{phase4eq}
 \end{equation}
 Note that Eq.~(\ref{phase4eq}) can be obtained from Eq.~(\ref{phase3eq}) simply by replacing the frame coherent state amplitude modulus squared $|\alpha_0|^2$ with the frame Fock state number $N$. However, in contrast to the leading order accumulated phase difference for the coherent state and classical oscillating frame, the phase difference for the Fock state frame depends explicitly
on the frame mass $M$. 
 
 As our second quantum frame example, consider a frame initially in a quadrature squeezed vacuum state:  $\hat{\rho}_{\mathrm{frame}}(0)=|\xi\rangle\langle \xi|=\hat{S}(\xi)|0\rangle\langle 0|\hat{S}^{\dagger}(\xi)$, where the squeeze operator is defined as $\hat{S}(\xi)=\exp\left[\frac{1}{2}\left(\xi^* \hat{a}^2-\xi \hat{a}^{\dagger 2}\right)\right]$~\cite{gerry}.  Equation~(\ref{interfere3eq}) then gives 
 \begin{eqnarray}
 \left\langle e^{i\hat{\phi}}\right\rangle&\approx &\frac{e^{i\frac{m}{8M}(\Omega-\omega)T}}{\sqrt{\cosh^2(r)-e^{i\frac{m}{4M}\Omega T} \sinh^2(r)}}\cr
 &\approx&\frac{e^{i\frac{m}{8M}(\Omega-\omega)T}}{\sqrt{1-i\frac{m}{4M}\Omega T \sinh^2(r)}},
 \label{squeezephaseeq}
 \end{eqnarray}
 where $\xi=re^{i\theta}$. Note that the interference term is suppressed for sufficiently large squeeze parameter $r$ such that 
 \begin{equation}
 \sinh^2(r)\approx e^{2 r}/4\gg 4 M/(m\Omega T)\gg 1. 
 \label{condition1eq}
 \end{equation}

 As our final example, we consider a frame initially in a thermal state:  $\hat{\rho}_{\mathrm{frame}}(0)=Z^{-1}\sum_{N=0}^{\infty} e^{-\beta\hbar\Omega (N+1/2)}|N\rangle\langle N|$.  In this case, Eq.~(\ref{interfere3eq}) gives
 \begin{equation}
 \left\langle e^{i\hat{\phi}}\right\rangle\approx e^{-i\frac{m\omega T}{8M}}\frac{\sinh(\beta\hbar\Omega/2)}{\sinh\left(\beta\hbar\Omega/2-im\Omega T/(8M)\right)}.
 \label{phase5eq}
 \end{equation}
Note that the interference is suppressed at sufficiently large temperatures such that 
\begin{equation}
(\hbar \Omega\beta)^{-1}\gg 4 M/(m\Omega T) \gg 1.
\label{condition2eq}
\end{equation}

The interference suppression conditions~(\ref{condition1eq}) and (\ref{condition2eq}) can be made more transparent by expressing in terms of the atom energy uncertainty arising from the corresponding frame energy uncertainty. For the example of the squeezed frame state, the frame energy uncertainty is $\Delta E_{\mathrm{frame}}=\hbar\Omega \sinh(2 r)/(\sqrt{2})\approx \hbar\Omega\sqrt{2} e^{2r}/4$ for $r\gg 1$, while for the thermal frame state, the frame energy uncertainty is $\Delta E_{\mathrm{frame}}\approx\beta^{-1}$ for $\beta^{-1}\gg\hbar\Omega$. Substituting into the respective conditions (\ref{condition1eq}) and (\ref{condition2eq}), we obtain the following common condition on the atom dwell time for the loss of quantum interference:
\begin{equation}
T\gg \frac{\hbar}{\frac{m}{M}\Delta E_{\mathrm{frame}}}\approx \frac{\hbar}{\Delta E_{\mathrm{atom}}},
\label{commonconditioneq}
\end{equation} 
where we have neglected numerical factors. Thus, loss of interference occurs when the atom dwell time in the vibrating wire frame  exceeds a dephasing timescale given by the atom's energy uncertainty. We may speculate that Eq.~(\ref{commonconditioneq}) is a general condition for dephasing, together with the requirement that the initial frame state is such that the probabilities $P_N=\langle N|\hat{\rho}_{\mathrm{frame}}(0)|N\rangle$ are broadly distributed in the Fock state number $N$, as is the case for the squeezed state with $r\gg 1$ and thermal state with $\beta^{-1}\gg\hbar\Omega$. 

While a quantum frame state with a sufficiently large energy uncertainty can lead to a suppression of atom interference, so too can a mixed, classical frame  state, as we have just seen for the squeezed vacuum and thermal state examples; it is not possible to qualitatively distinguish between quantum and classical frame states simply by measuring the atom wave  interference for the  two-path interferometer model that we are considering.  Furthermore, while it is unlikely that we would be able to realize anytime soon such substantial squeezing~\cite{wollman15} as required by~(\ref{condition1eq}) for interference suppression,  mesoscale mechanical resonators viewed as quantum non-inertial reference frames are nevertheless of theoretical interest for the unusual insights that are gained.      

 Just as for the classical frame interference, it is interesting to determine whether the quantum frame interference follows from a more fundamental ``twin paradox" description, where the proper time of the atom wave packet traversing the vibrating wire frame now develops a quantum uncertainty as a result of the frame being in a quantum state. Formally, the averaged interference term in (\ref{interfere2eq}) might be expressed as [cf. Eq.~(\ref{phase2eq})]:
 \begin{equation}
  \left\langle e^{i\hat{\phi}}\right\rangle=\left\langle {\cal{T}} e^{-i\frac{m c^2}{\hbar}\left[\int_0^T dt (d\hat{\tau}_R/dt)-T\right]}\right\rangle,
  \label{phase6eq}
  \end{equation}
where ${\cal{T}}$ denotes lab coordinate time ordering and $\hat{\tau}_R$ denotes a quantum proper time operator. A path integral formulation~\cite{hartle86,mueller13}  of~(\ref{phase6eq}) may be possible, with the class of interfering world line paths summed over restricted by the requirement that the formulation must reproduce the quantum frame interference expression~(\ref{interfere3eq}) in the non-relativistic limit.  In this respect, treating mesoscopic mechanical resonators as quantum reference frames may yield, via the equivalence principle, possible insights concerning the nature of quantum gravity at low energies~\cite{aharonov84}, in particular the effect of quantum fluctuating space-time on matter wave interference~\cite{blencowe13,anastopoulos13}.  

\section{possible realizations}
\label{sec:realization}

We now outline possible methods for realizing the mesoscopic, vibrating wire interferometer. As was discussed, to realize a phase difference $\phi\sim 1$ with an atom and an acceleration of $10^{8}~g$, one needs a wire frame dwell time of $\sim 1~\mu{\mathrm{s}}$.  For a nanomechanical resonator with a length of approximately $1~\mu{\mathrm{m}}$, this requires an atom velocity of $1~{\mathrm{ms}}^{-1}$ or less. Alternatively,  electrons, given their much smaller mass, would require a dwell time approximately 1000 times longer [see Eq.~(\ref{phase1eq})], and hence a velocity of $<10^{-3}~{\mathrm{ms}}^{-1}$ to realize a substantial phase difference in a similar length. As a consequence, atoms appear to be much more favorable for a possible experiment.  

\begin{figure}[htbm]
\centering
{\includegraphics[width=7.0cm]{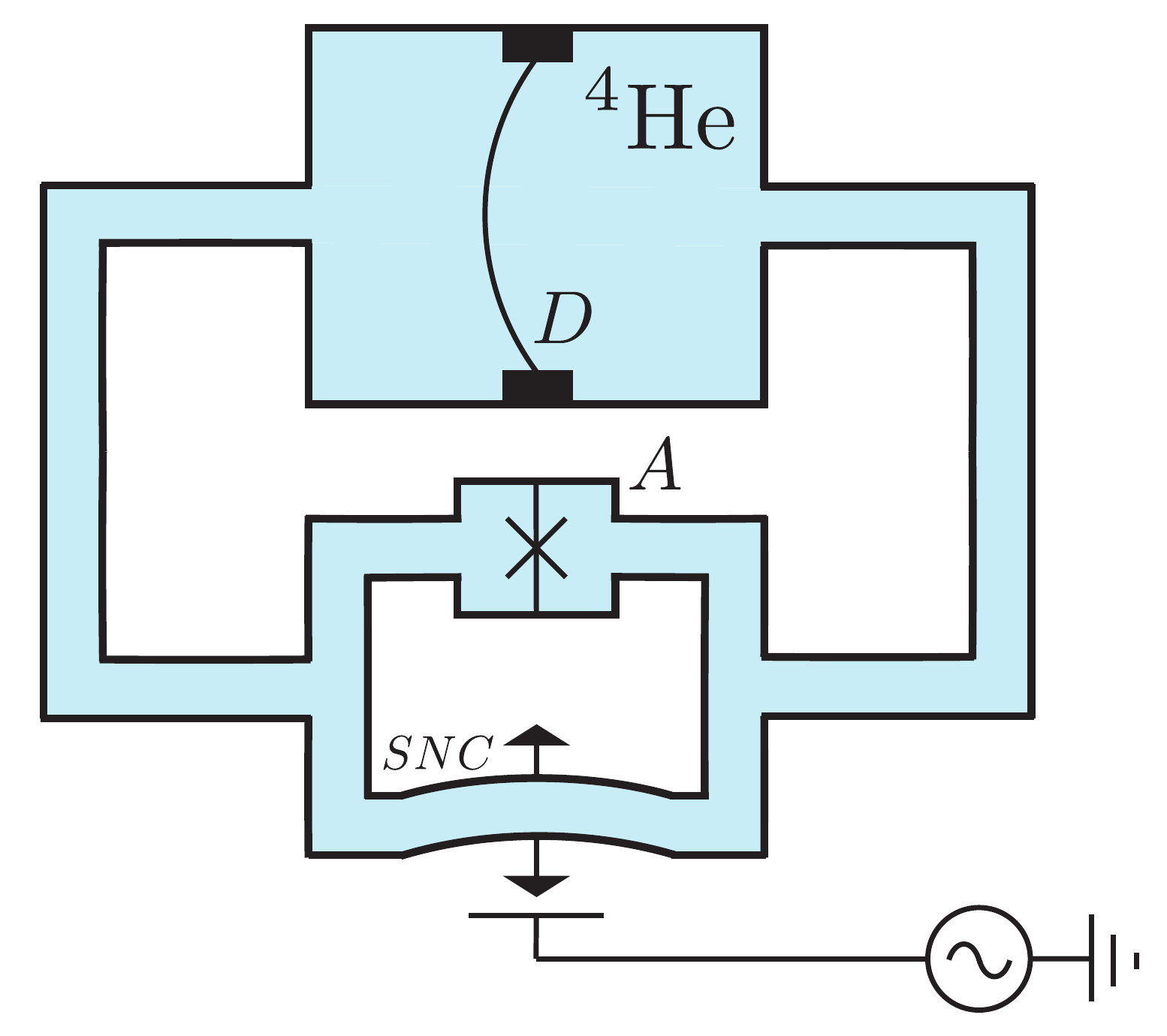}
}
\caption{Superfluid interferometer setup: $D$ is the diaphragm which is used to both impose a chemical potential through hydrostatic pressure and measure the resulting superflow, $A$ is the $100~{\mathrm{nm}}$ diameter aperture used to monitor the quantum phase difference, $SNC$ is the suspended nanochannel filled with superfluid ${}^4{\mathrm{He}}$ moving through the channel at $1~{\mathrm{ms^{-1}}}$ and accelerating transversely at $10^8~g$,  driven into motion with an external drive field.}
\label{fig:superfluid}
\end{figure}

Fig.~\ref{fig:superfluid} shows a scheme for a possible superfluid interferometer device used to detect quantum phase differences in superfluid helium~\cite{schwab1997,schwab1998,sato12}. One arm  of the interferometer is interrupted by a $100~{\mathrm{nm}}$ diameter aperture in a thin silicon nitride membrane, and the other arm a suspended nanochannel mechanical resonator~\cite{Olcum2014} or nanopipe~\cite{Whitby2007} with diameter-to-length dimensions $250~{\mathrm{nm}}\times 1~\mu{\mathrm{m}}$. The quantum phase difference at critical velocities through sub-micron apertures is typically $\sim 2\pi\cdot 10$, which yields a superfluid velocity in the suspended channel of $\sim 1~{\mathrm{ms}}^{-1}$. The quantum phase that is generated through the large acceleration of the suspended channel will produce a mass current through the aperture, which modifies the apparent critical velocity that is measured with an external diaphragm.   The nanochannel diameter should be sufficiently narrow to avoid the excitation of transverse acoustic modes at the drive frequency: the speed of first sound in superfluid $c_{\mathrm{He}}=240~{\mathrm{ms}}^{-1}$, yielding an acoustic wavelength  $\lambda = c_{\mathrm{He}}/f=2.4~\mu{\mathrm{m}}$ for $f=100~{\mathrm{MHz}}$.  It appears possible to integrate a circuit with these elements on the surface of a chip~\cite{schwab1996}. Other possible realizations could involve cold atomic clouds or Bose-Einstein condensates steered on the surface of atom-chips~\cite{schumm2005}, or guided through hollow fiber-optic atomic and optical waveguides~\cite{Bajcsy2011}. 
 
\section{conclusion}
In this paper, we have explored several consequences of mesoscopic mechanical resonators viewed as classical and  quantum non-inertial reference frames. We considered a simple model, two-path atom interferometer set-up, where one of the paths is furnished by a vibrating wire atom guide. For the example of a hundred megahertz vibrating, micron scale wire frame, nanometer amplitude wire displacements  can induce significant  phase shifts between  the interfering atom quantum wave function components for the two paths. We also showed that a vibrating wire with sufficiently large quantum or classical energy uncertainty can result in the loss of quantum interference between the two atom wave function components. While we suggested a possible approach towards realizing such interferometers using a superfluid helium  set-up, it is unlikely that significant quantum non-inertial frame signatures in atom wave interference can be demonstrated anytime soon. Nevertheless, viewing mesoscopic vibrating wires as quantum reference frames is of fundamental theoretical interest. By making the correspondence with classical frames, we raised the possibility of  a quantum proper time formulation of the two path interference. Such a formulation may provide insights concerning the effects of quantum fluctuating space-time on matter wave interference.      

\section*{acknowledgements}
We  thank Prof. S.  A. Werner for  conversations that inspired this work. We acknowledge funding provided by the Foundational Questions Institute (FQXi), the National Science Foundation under Grant No.  DMR-1104790 (MPB), and the Institute for Quantum Information and Matter, an NSF Physics Frontiers Center with support from the Gordon and Betty Moore Foundation through Grant GBMF1250 (KCS). 

\appendix*
\section{Derivation of the quantum frame accumulated phase difference} 
In this section, we  derive the expression (7) for the averaged accumulated phase difference $\left\langle e^{i\hat{\phi}}\right\rangle$ between atom wave components traversing the two-path interferometer, where a segment of the right path involves a non-inertial,  vibrating wire quantum frame.  The initial state of the atom-frame system is
\begin{eqnarray}
\frac{1}{2}&&\left(|\psi_L(0)\rangle\langle\psi_L(0)|+|\psi_R(0)\rangle\langle\psi_R(0)|+|\psi_L(0)\rangle\langle\psi_R(0)|\right.\cr
&&\left.+|\psi_R(0)\rangle\langle\psi_L(0)|\right)\otimes\hat{\rho}_{\mathrm{frame}}(0),
\label{initstateeq}
\end{eqnarray}
where $\psi_L(x,0)=\psi_R(x,0)=\left(\frac{m\omega}{\pi\hbar}\right)^{1/4}\exp\left(-\frac{m\omega x^2}{2\hbar}\right)$, i.e., the atom enters the interferometer arms initially in its transverse ground state and the frame is initially in some (possibly mixed) state $\hat{\rho}_{\mathrm{frame}}(0)$.
After the atom has exited the quantum frame at time $T$, the atom-frame state is
\begin{eqnarray}
&&\frac{1}{2}\hat{U}_L(T)|\psi_L(0)\rangle\langle\psi_L(0)|\otimes\hat{\rho}_{\mathrm{frame}}(0)\hat{U}_L^{\dagger}(T)\cr
+&&\frac{1}{2}\hat{U}_R(T)|\psi_R(0)\rangle\langle\psi_R(0)|\otimes\hat{\rho}_{\mathrm{frame}}(0)\hat{U}_R^{\dagger}(T)\cr
+&&\frac{1}{2}\hat{U}_L(T)|\psi_L(0)\rangle\langle\psi_R(0)|\otimes\hat{\rho}_{\mathrm{frame}}(0)\hat{U}_R^{\dagger}(T)\cr
+&&\frac{1}{2}\hat{U}_R(T)|\psi_R(0)\rangle\langle\psi_L(0)|\otimes\hat{\rho}_{\mathrm{frame}}(0)\hat{U}_L^{\dagger}(T),
\label{finalstateeq}
\end{eqnarray}
where the unitary operators $\hat{U}_{L(R)}(T)$ implement the atom-frame evolution over the time interval $T$ when the atom is either attached to the frame, i.e., traversing the right arm ($R$), or not attached to the frame, i.e., traversing the left arm $(L)$.

The interference between left and right path wave components at $t=T$ is (cf. the classical frame interference Eq.~(2)):
\begin{eqnarray}
{\mathrm{Tr}}_{\mathrm{frame}}&&\left[\langle x|\hat{U}_R(T)|\psi_R(0)\rangle\langle\psi_L(0)|\otimes\hat{\rho}_{\mathrm{frame}}(0)\hat{U}_L^{\dagger}(T)|x\rangle\right]\cr
&&=\sqrt{\frac{m\omega}{\pi\hbar}}e^{-\frac{m\omega x^2}{\hbar}} \left\langle e^{i\hat{\phi}}\right\rangle.
\label{interfereq}
\end{eqnarray}
Implementing the trace in terms of the frame position states $|X\rangle$ and inserting the frame coherent state resolution of unity,  $1=\frac{1}{\pi}\int d^2\alpha |\alpha\rangle\langle\alpha|$, the interference~(\ref{interfereq}) can be written as
 \begin{eqnarray}
 \frac{1}{\pi^2}&&\int d^2\alpha\int d^2\alpha'\int_{-\infty}^{+\infty} dX\langle x|\langle X|\hat{U}_R(T)|\psi_R(0)\rangle|\alpha\rangle\cr
 &&\times\langle\alpha|\hat{\rho}_{\mathrm{frame}}(0)|\alpha'\rangle\langle\psi_L(0)|\langle\alpha'|\hat{U}^{\dagger}_L(T)|x\rangle|X\rangle.
 \label{interfer2eq}
 \end{eqnarray}
Equation~(\ref{interfer2eq}) expresses the interference term for an arbitrary initial frame state $\hat{\rho}_{\mathrm{frame}}(0)$ in terms of the  evolution of initial atom-frame coherent states $\Phi_{\alpha R}(x,X,T)\equiv\langle x|\langle X|\hat{U}_R(T)|\psi_R(0)\rangle|\alpha\rangle$ and $\Phi_{\alpha L}(x,X,T)\equiv\langle x|\langle X|\hat{U}_L(T)|\psi_L(0)\rangle|\alpha\rangle$, where the frame coherent state wave function is 
\begin{eqnarray}
&&\Psi_{\alpha}(X)\equiv\langle X|\alpha\rangle=\left(\frac{M\Omega}{\pi\hbar}\right)^{1/4}\exp\left[-\frac{1}{4}\left(\frac{X}{X_{\mathrm{zp}}}\right)^2\right.\cr
&&\left. +\alpha\left(\frac{X}{X_{\mathrm{zp}}}\right)-\alpha \operatorname{Re} (\alpha)\right],
\label{coherentstateeq}
\end{eqnarray}   
with $X_{\mathrm{zp}}=\sqrt{\frac{\hbar}{2M\Omega}}$  the frame zero point uncertainty. The wave function $\Phi_{\alpha R}$ describing the attached atom-frame composite system is a solution to the Schr\"{o}dinger equation~(5), reproduced here:
\begin{eqnarray}
i\hbar\frac{\partial\Phi_{\alpha R}}{\partial t}&=&-\frac{\hbar^2}{2 m} \frac{\partial^2\Phi_{\alpha R}}{\partial x^2}+\frac{1}{2}m\omega^2\left[x-X\sin\left(\pi t/T\right)\right]^2 \Phi_{\alpha R}\cr
&&-\frac{\hbar^2}{2 M} \frac{\partial^2\Phi_{\alpha R}}{\partial X^2}+\frac{1}{2}M\Omega^2 X^2 \Phi_{\alpha R}.
\label{schReq}
\end{eqnarray}
On the other hand, the wave function $\Phi_{\alpha L}$, describing the atom-frame composite system with the atom not attached to the frame, is a solution to the following Schr\"{o}dinger equation:
\begin{eqnarray}
i\hbar\frac{\partial\Phi_{\alpha L}}{\partial t}&=&-\frac{\hbar^2}{2 m} \frac{\partial^2\Phi_{\alpha L}}{\partial x^2}+\frac{1}{2}m\omega^2 x^2 \Phi_{\alpha L}\cr
&&-\frac{\hbar^2}{2 M} \frac{\partial^2\Phi_{\alpha L}}{\partial X^2}+\frac{1}{2}M\Omega^2 X^2 \Phi_{\alpha L}, \label{schLeq}
\end{eqnarray}
which simply describes two decoupled harmonic oscillators. It is convenient to work in terms of dimensionless coordinates $\tilde{x}=x/x_{\mathrm{zp}}$, with $x_{\mathrm{zp}}=\sqrt{\frac{\hbar}{2m\omega}}$,  $\tilde{X}=X\sqrt{2M\omega/\hbar}$, and $\tau=\omega t$. The Schr\"{o}dinger equations~(\ref{schReq}) and (\ref{schLeq}) then become respectively:
\begin{equation}
i\frac{\partial\Phi_{\alpha R}}{\partial \tau}=\left\{-\frac{\partial^2}{\partial \tilde{x}^2}-\frac{\partial^2}{\partial \tilde{X}^2}+\frac{1}{4}\left(\begin{array}{cc} \tilde{x} & \tilde{X}\end{array}\right){\mathcal{V}}(\tau)\left(\begin{array}{c}\tilde{x}\\ \tilde{X}\end{array}\right)\right\}\Phi_{\alpha R}
\label{schR2eq}
\end{equation}
and
\begin{equation}
i\frac{\partial\Phi_{\alpha L}}{\partial \tau}=\left\{-\frac{\partial^2}{\partial \tilde{x}^2} +\frac{1}{4} \tilde{x}^2-\frac{\partial^2}{\partial \tilde{X}^2}+\frac{1}{4} \left(\frac{\Omega}{\omega}\right)^2 \tilde{X}^2 \right\}\Phi_{\alpha L},
\label{schL2eq}
\end{equation}
where the coupled potential energy term in the attached atom-frame Schr\"{o}dinger equation~(\ref{schR2eq}) has been put in matrix form to indicate more clearly the possibility to diagonalize the matrix (i.e., transform to decoupled coordinates):
\begin{equation}
{\mathcal{V}}(\tau)= \left(\begin{array}{cc} 1& -\sqrt{\frac{m}{M}}\sin\left(\frac{\pi\tau}{\omega T}\right)\\
-\sqrt{\frac{m}{M}}\sin\left(\frac{\pi\tau}{\omega T}\right)&~~\left(\frac{\Omega}{\omega}\right)^2+\frac{m}{M}\sin^2\left(\frac{\pi\tau}{\omega T}\right)\end{array}\right).
\label{matrixeq}
\end{equation}
In particular, noting that ${\mathcal{V}}(\tau)$ is symmetric and hence has real eigenvalues $\lambda_{\pm}(\tau)$ with associated orthonormal eigenvectors ${\mathbf{v}}_{\pm}(\tau)$:
\begin{equation}
{\mathcal{V}}(\tau){\mathbf{v}}_{\pm}(\tau)=\lambda_{\pm}(\tau){\mathbf{v}}_{\pm}(\tau),
\label{eigenvalueeq}
\end{equation}
we have that
\begin{equation}
{\mathcal{S}}^T(\tau) {\mathcal{V}}(\tau){\mathcal{S}}(\tau)=\left(\begin{array}{cc} \lambda_+ (\tau)& 0\\
0&\lambda_- (\tau)\end{array}\right),
\label{transformmatrixeq}
\end{equation}
where the transformation matrix is ${\mathcal{S}}(\tau)=\left({\mathbf{v}}_+(\tau),{\mathbf{v}}_-(\tau)\right)$.
Introducing  coordinates $\xi_{\pm}$:
\begin{equation}
\left(\begin{array}{c}\tilde{x}\\ \tilde{X}\end{array}\right)={\mathcal{S}}(\tau)\left(\begin{array}{c}\xi_+\\ \xi_-\end{array}\right),
\label{transformeq}
\end{equation}
Schr\"{o}dinger equation~(\ref{schR2eq}) becomes
\begin{equation}
i\frac{\partial\Phi_{\alpha R}}{\partial\tau}=\left\{-\frac{\partial^2}{\partial\xi_+^2}+\frac{1}{4}\lambda_+(\tau)\xi_+^2 -\frac{\partial^2}{\partial\xi_-^2}+\frac{1}{4}\lambda_-(\tau)\xi_-^2\right\}\Phi_{\alpha R}.
\label{transfschReq}
\end{equation}

From Eq.~(\ref{coherentstateeq}), the initial, attached atom-frame composite coherent state is
\begin{eqnarray}
&&\Phi_{\alpha R}\left(\xi_+,\xi_-,0\right)=\Phi_{\alpha R}\left(\tilde{x},\tilde{X},0\right)=\left(\frac{m\omega}{\pi\hbar}\right)^{1/4} \exp\left(-\frac{\tilde{x}^2}{4}\right)\cr
&&\times\left(\frac{M\Omega}{\pi\hbar}\right)^{1/4}\exp\left(-\frac{\Omega}{4\omega}\tilde{X}^2 +\alpha\sqrt{\frac{\Omega}{\omega}}\tilde{X}-\alpha\operatorname{Re} (\alpha)\right),\cr
&&
\label{initstate2eq}
\end{eqnarray}
where the first equality follows from the fact that ${\mathcal{S}} (\tau)$ is simply the identity matrix at $\tau=0$. Given the form of Eq.~(\ref{transfschReq}) and the fact that the initial state~(\ref{initstate2eq}) is a product state, we have for the final state:
\begin{eqnarray}
\Phi_{\alpha R}&\left(\tilde{x},\tilde{X},\omega T\right)= \Phi_{\alpha R}\left(\xi_+,\xi_-,\omega T\right)\cr
&=\psi_R\left(\xi_+,\omega T\right)\Psi_{\alpha R}\left(\xi_-,\omega T\right),
\label{finalstate2eq}
\end{eqnarray}
where the first equality follows from the fact that ${\mathcal{S}} (\tau)$ is simply the identity matrix at $\tau=\omega T$, and where $\psi_R$ and $\Psi_{\alpha R}$ are solutions to the decoupled Schr\"{o}dinger equations:
\begin{equation}
i\frac{\partial\psi_{R}}{\partial\tau}=\left\{-\frac{\partial^2}{\partial\xi_+^2}+\frac{1}{4}\lambda_+(\tau)\xi_+^2\right\}\psi_{R}
\label{decouple1schReq}
\end{equation}
and
\begin{equation}
i\frac{\partial\Psi_{\alpha R}}{\partial\tau}=\left\{-\frac{\partial^2}{\partial\xi_-^2}+\frac{1}{4}\lambda_-(\tau)\xi_-^2\right\}\Psi_{\alpha R},
\label{decouple2schReq}
\end{equation}
with the respective initial conditions following from Eq.~(\ref{initstate2eq}):
\begin{equation}
\psi_{R}(\xi_+,0)=\psi_{R}(\tilde{x},0)=\left(\frac{m\omega}{\pi\hbar}\right)^{1/4} \exp\left(-\frac{\tilde{x}^2}{4}\right)
\label{initpluseq}
\end{equation}
and
\begin{eqnarray}
\Psi_{\alpha R}(\xi_-,0)&=&\Psi_{\alpha R}(\tilde{X},0)=\left(\frac{M\Omega}{\pi\hbar}\right)^{1/4}\exp\left(-\frac{\Omega}{4\omega}\tilde{X}^2\right. \cr
&+&\left.\alpha\sqrt{\frac{\Omega}{\omega}}\tilde{X}-\alpha\operatorname{Re} (\alpha)\right).
\label{initminuseq}
\end{eqnarray}

The eigenvalue solutions to Eq.~(\ref{eigenvalueeq}) are
\begin{eqnarray}
&&\lambda_{\pm}(\tau)=\frac{1}{2}\left(1+\left(\frac{\Omega}{\omega}\right)^2+\frac{m}{M}\sin^2\left(\frac{\pi\tau}{\omega T}\right)\right)\cr
&&\pm\frac{1}{2}\sqrt{\left(1+\left(\frac{\Omega}{\omega}\right)^2+\frac{m}{M}\sin^2\left(\frac{\pi\tau}{\omega T}\right)\right)^2-4\left(\frac{\Omega}{\omega}\right)^2},\cr
&&\label{eigenvalues}
\end{eqnarray}
which under the conditions $m \lll M$ and $\Omega\ll\omega$,  can be approximated as
\begin{equation}
\lambda_+(\tau)\approx 1+\frac{m}{M}\sin^2\left(\frac{\pi\tau}{\omega T}\right)
\label{lambdaplusapproxeq}
\end{equation}
and
\begin{equation}
\lambda_-(\tau)\approx\left(\frac{\Omega}{\omega}\right)^2\left(1-\frac{m}{M}\sin^2\left(\frac{\pi\tau}{\omega T}\right)\right).
\label{lambdaminusapproxeq}
\end{equation} 
Solving the decoupled Schr\"{o}dinger equations~(\ref{decouple1schReq}), (\ref{decouple2schReq}) with approximate eigenvalues~(\ref{lambdaplusapproxeq}), (\ref{lambdaminusapproxeq}), and initial conditions~(\ref{initpluseq}), (\ref{initminuseq}), it is convenient to assume that the solutions are Gaussian: 
\begin{equation}
\psi_R(\xi_+,\tau)=\left(\frac{m\omega}{\pi\hbar}\right)^{1/4}\exp\left[a_+ (\tau)\xi_+^2 +b_+(\tau)\xi_+ +c_+(\tau)\right]
\label{gaussianpluseq}
\end{equation}
and
\begin{eqnarray}
\Psi_{\alpha R} (\xi_-,\tau)=\left(\frac{M\Omega}{\pi\hbar}\right)^{1/4}\exp&& \left[a_- (\tau)\xi_-^2 +b_-(\tau)\xi_-\right. \cr
&&\left.+c_-(\tau)\right].
\label{gaussianminuseq}
\end{eqnarray}
Substituting Eqs.~(\ref{gaussianpluseq}) and (\ref{gaussianminuseq}) into their respective Schr\"{o}dinger equations~(\ref{decouple1schReq}) and (\ref{decouple2schReq}), we obtain the following equations for the time dependent coefficients:
\begin{eqnarray}
&&\frac{d a_{\pm}}{d\tau}-4 i a^2_{\pm} +\frac{i}{4}\lambda_{\pm}(\tau)=0\cr
&&\frac{d b_{\pm}}{d\tau}-4 i b_{\pm} a_{\pm}=0\cr
&&\frac{d c_{\pm}}{d\tau}-2 i a_{\pm} -i b_{\pm}^2=0.
\label{coeffeq}
\end{eqnarray}
The approximate solutions to these coefficients at $\tau=\omega T$ are:
\begin{eqnarray}
&&a_+ (\omega T)\approx -\frac{1}{4}\cr
&&b_+ (\omega T)\approx 0\cr
&&c_+ (\omega T)\approx -i\frac{\omega T}{2}\left(1+\frac{m}{4M}\right)
\label{coeffplussolneq}
\end{eqnarray}
and
\begin{eqnarray}
&&a_- (\omega T)\approx -\frac{\Omega}{4\omega}\cr
&& b_- (\omega T)\approx \alpha \sqrt{\frac{\Omega}{\omega}}\exp\left[-i\Omega T\left(1-\frac{m}{4M}\right)\right]\cr
&&c_- (\omega T)\approx -i\frac{\Omega T}{2}\left(1-\frac{m}{4M}\right)\cr
&&+\frac{\alpha^2}{2} \left(1-e^{-2i\Omega T\left(1-m/(4M)\right)}\right)-\alpha\operatorname{Re} (\alpha).
\label{coeffminussolneq}
\end{eqnarray}
Substituting the solutions~(\ref{coeffplussolneq}) and (\ref{coeffminussolneq}) into Eq.~(\ref{finalstate2eq}), we obtain for the evolution of the attached atom-frame coherent states:
\begin{eqnarray}
&\Phi_{\alpha R}& (x, X, T)\approx\left(\frac{m\omega}{\pi\hbar}\right)^{1/4}\exp\left(-\frac{m\omega x^2}{2\hbar}\right) e^{-i\frac{\omega T}{2}\left(1+\frac{m}{4M}\right)}\cr
&&\times\langle X|\alpha e^{-i\Omega T\left(1-m/(4M)\right)}\rangle e^{-i\frac{\Omega T}{2}\left(1-\frac{m}{4M}\right)}.
\label{phirighteq}
\end{eqnarray}
An analogous but considerably more straightforward analysis gives the evolution of the atom-frame coherent states when the atom traverses the left path (i.e., is not attached to the frame):
\begin{eqnarray}
&\Phi_{\alpha L} &(x, X, T)=\left(\frac{m\omega}{\pi\hbar}\right)^{1/4}\exp\left(-\frac{m\omega x^2}{2\hbar}\right) e^{-i\frac{\omega T}{2}}\cr
&&\times\langle X|\alpha e^{-i\Omega T}\rangle e^{-i\frac{\Omega T}{2}}.
\label{philefteq}
\end{eqnarray}
Substituting the solutions (\ref{phirighteq}) and (\ref{philefteq}) into the interference expression~(\ref{interfer2eq}), performing the trace over the frame subsystem (i.e., integration with respect to the $X$ coordinate), we obtain:
\begin{eqnarray}
\left\langle e^{i\hat{\phi} }\right\rangle\approx&\frac{1}{\pi^2}&\int d^2\alpha\int d^2\alpha' \langle\alpha'e^{-i\Omega T}|\alpha e^{-i\Omega T\left(1-m/(4M)\right)}\rangle\cr
&&\times\langle\alpha|\hat{\rho}_{\mathrm{frame}}(0)|\alpha'\rangle e^{i\frac{m}{8M}(\Omega-\omega)T}
\label{interfer3eq}
\end{eqnarray}
From the equality $\langle\alpha'e^{-i\Omega T}|\alpha e^{-i\Omega T\left(1-m/(4M)\right)}\rangle=\langle\alpha'|\alpha e^{im\Omega T/(4M)}\rangle$, the $\alpha'$ integral can be carried out and we finally obtain Eq.~(7) for the averaged accumulated phase difference between the left and right path atom waves:
\begin{equation}
 \left\langle e^{i\hat{\phi}}\right\rangle\approx\frac{1}{\pi}\int d^2\alpha\langle\alpha|\hat{\rho}_{\mathrm{frame}}(0)|\alpha e^{im\Omega T/(4M)}\rangle e^{i\frac{m}{8 M}(\Omega-\omega)T}.
 \label{interferefinaleq}
 \end{equation}

\end{document}